# Superconducting Magnet with the Reduced Barrel Yoke for the Hadron Future Circular Collider

V. I. Klyukhin, A. Ball, C. Berriaud, B. Curé, A. Dudarev, A. Gaddi, H. Gerwig, A. Hervé, M. Mentink, G. Rolando, H. F. Pais Da Silva, U. Wagner, and H. H. J. ten Kate

*Abstract*– The conceptual design study of a hadron Future Circular Collider (FCC-hh) with a center-of-mass energy of the order of 100 TeV in a new tunnel of 80-100 km circumference assumes the determination of the basic requirements for its detectors. A superconducting solenoid magnet of 12 m diameter inner bore with the central magnetic flux density of 6 T is proposed for a FCC-hh experimental setup. The coil of 24.518 m long has seven 3.5 m long modules included into one cryostat. The steel yoke with a mass of 21 kt consists of two barrel layers of 0.5 m radial thickness, and 0.7 m thick nose disk, four 0.6 m thick end-cap disks, and three 0.8 m thick muon toroid disks each side. The outer diameter of the yoke is 17.7 m; the length without the forward muon toroids is 33 m. The air gaps between the end-cap disks provide the installation of the muon chambers up to the pseudorapidity of ±3.5. The conventional forward muon spectrometer provides the measuring of the muon momenta in the pseudorapidity region from ±2.7 to ±4.6. The magnet modeled with Cobham's program TOSCA. The total Ampere-turns in the superconducting solenoid coil are 127.25 MA-turns. The stored energy is 43.3 GJ. The axial force onto each end-cap is 480 MN. The stray field at the radius of 50 m off the coil axis is 14.1 mT and 5.4 mT at the radius of 100 m. All other parameters presented and discussed.

## I. INTRODUCTION

THE hadron Future Circular Collider (FCC-hh) [1] with a center-of-mass energy of the order of 100 TeV assumed to be constructed in a new tunnel of 80-100 km circumference, requires to use in the experimental setups the superconducting solenoid coils with a free bore of 12 m in diameter and with the central magnetic flux density of 6 T. The future progress in the tracking detectors will allow measuring the momenta of the prompt muons inside the inner tracker, if the muon system will indicate the charged tracks are really the muons. In this case, the barrel part of the external muon system could be simplified using rather thin steel yoke with the main purpose to eliminate the low momentum muons arising from the hadron decays in flight, and the punch through hadrons to ensure the prompt muon identification. The magnetic flux density bending component integral of about 3.5 T·m will be enough to perform this task.

The physics requirements assume the location of the major sub-detectors inside the superconducting coil. The sub-detectors are the inner tracker of 5 m outer diameter with the length of 16 m, the electromagnetic calorimeter with the outer diameter of 7.2 m and the length of 18.2 m, and the hadronic calorimeter with the outer diameter of 12 m and the length of at least of 23 m.

## II. MODEL DESCRIPTION

Fig. 1 presents a three-dimensional (3-D) FCC-hh detector magnetic system model based on the CMS magnet experience [2], [3], and developed and calculated with Cobham's program TOSCA [4].

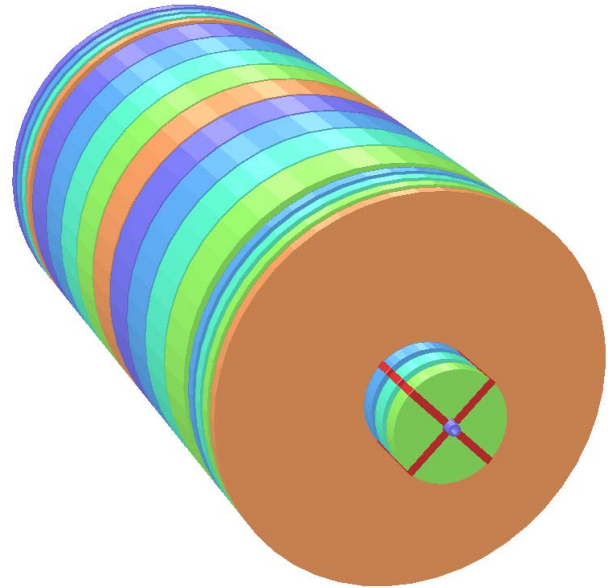

Fig. 1. 3-D model of the FCC-hh detector magnetic system.

### A. Superconducting Coil

The magnetic system includes the superconducting coil with 127.25 MA-turns that creates the magnetic flux density of 5.9906 T in the center. The coil inner diameter of 6.19 m at the room temperature and the length of 24.518 m keep the same the diameter to length ratio as is used in the CMS magnet [2]. The coil consists of seven modules of 3.5 m long with 3 mm thick insulation between the modules.

To wind 6 layers of the coil inside the copper quench back cylinder of 0.1 m thick the Cu-stabilized conductor with the cross-section of 22×68 mm² and the NbTi superconducting insert of 2.34×20.63 mm² could used. With the thickness of the insulation around the conductor of 0.5 mm, the additional insulation between six coil layers of 0.4 mm, and the insulation at the inner and outer radii of 1 mm, the coil radial thickness without the quench back cylinder is 0.418 m, and the conductor mass is not less than 3418 t.

The total number of turns is 6342 and the current corresponding to the central magnetic flux density of 5.9906 T is 20065 A. The stored energy in the coil of 43.3 GJ gives the ratio of the stored energy to the coil mass of 12.66 kJ/kg that is about the CMS value of this ratio [2], [3].

The axial pressure in the coil middle plane is 68.47 MPa; the average radial pressure is 14.35±0.79 MPa.

*B. The Magnet Yoke and the Muon Toroids*

Nine steel wheels of 2.65 m width form two 0.5 m thick barrel layers separated by the radial distance of 0.35 m as shown in Figs. 2 and 3. This gives the integral of the magnetic flux density bending component in the coil middle plane of 2.675 T·m at the radial distance from 7.15 to 8.85 m, the outer radius of the barrel yoke.

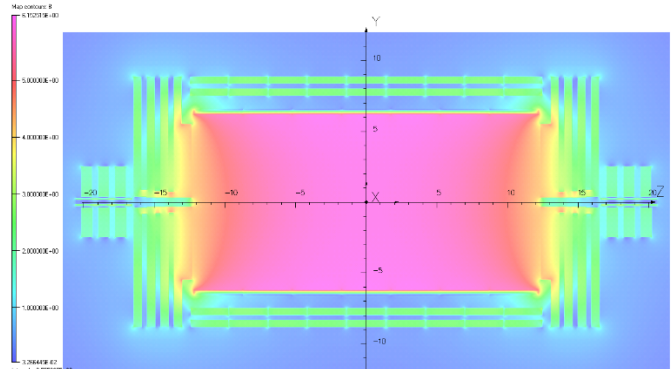

Fig. 2. Magnetic flux density distribution in the vertical plane. The color magnetic field map plotted with the cell size of 0.05 m has the width of 43 m and the height of 24 m. The color scale unit is 1 T. The minimum and maximum magnetic flux density values are 0.0327 and 6.1525 T.

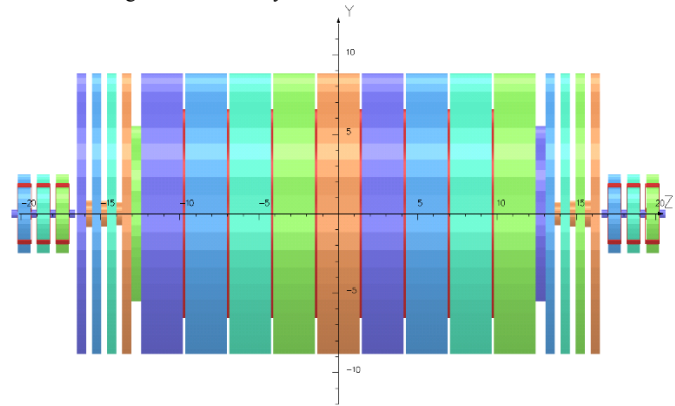

Fig. 3. The coil in red, the nine barrel wheels of 2.65 m width each, the two nose disks of 11 m diameter each, the eight end-cap disks of 17.7 m diameter each, the end-cap rings between the disks, and the six muon toroids of 5 m diameter with the conventional coils in red and the protection tubes inside.

To extend the homogenous region of the magnetic flux inside the coil, the steel nose disks with a thickness of 0.7 m and the outer diameter of 11 m are located on the both sides of the coil. The distance available between the nose disks for the sub-detectors is 24.9 m.

Four steel end-cap disks of 0.6 m thick connected by the steel rings of 0.35 m thick follow the nose disks at each coil side. The air gaps of 0.35 m between the end-cap disks allow to install the muon chambers covered the pseudorapidity [3] region of ±3.5.

The steel rings between the end-cap disks overshadow the pseudorapidity region from ±3.6 to ±4.5, and measuring the muon momenta with the end-cap muon chambers in this region is not possible.

Two forward muon spectrometers follow the end-cap disks at both sides of the yoke. Each spectrometer consists of three steel toroid disks of 0.8 m thick with the inner diameter of 0.732 m and the outer diameter of 5 m. There are four conventional copper coils at each toroid to magnetize steel with the current of 907.6 A. Each coil consists of 34 turns of 17.5×17.5 mm² copper conductor wound in two layers. The hole of 10 mm diameter in the conductor cross-section serves for water-cooling of the coils.

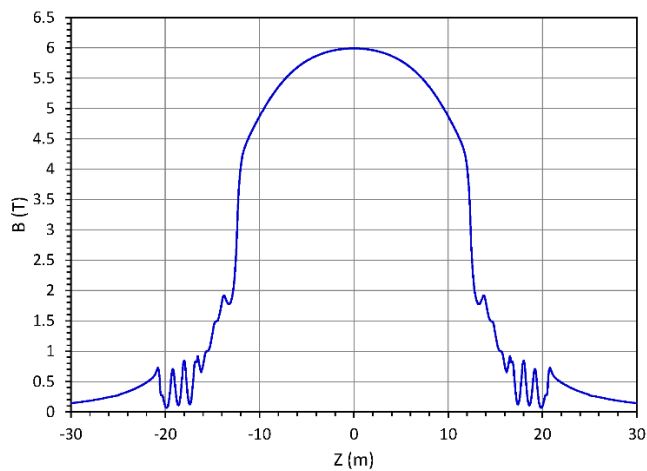

Fig. 4. Magnetic flux density variation along the coil axis.

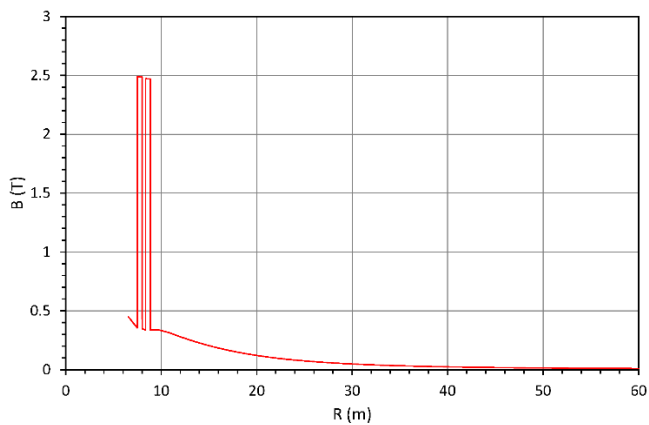

Fig. 5. Magnetic flux density out of the coil in the coil middle plane vs. radius.

The tubes of steel with the inner diameter of 0.3 m and the outer diameter of 0.54 m keeps the toroids in the positions providing the gaps of 0.4 m between the disks.

The total mass of the steel yoke is 20.74 kt, the outer diameter is 17.7 m, and the length included both forward muon spectrometers is 41.2 m.

### III. Magnet Parameters

In Fig. 2, the magnetic flux density distribution displayed in the vertical plane up to the radius of 12 m; Fig. 4 shows the magnetic flux density variation along the coil axis.

Fig. 5 presents the magnetic stray field variation vs. radius in the coil middle plane. The stray magnetic flux density drops to 14.1 mT at the radius of 50 m off the coil axis and to 5.4 mT at the radius of 100 m.

The axial force to each end-cap is 480 MN including the axial force to each forward muon spectrometer of 7.2 MN, and the maximum axial force to the barrel wheel is 46.8 MN.

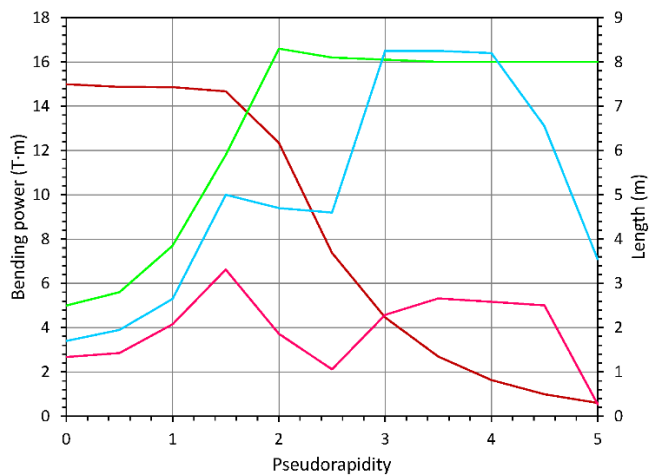

Fig. 6. Magnetic flux density bending component integrals (left scale) and the length of the charged particle trajectory (right scale) in the inner tracker (dark red and green curves), and in the muon system (pink and the light blue curves) vs. the pseudorapidity.

Fig. 6 displays the integrals of the magnetic flux density bending component orthogonal to the charged particle trajectory vs. pseudorapidity inside the inner tracker of 5 m diameter and 16 m length, and through the muon system. This plot shows that measuring the momenta of the charged particles in the inner tracker at the pseudorapidity greater than ±3.5 is rather difficult. The muon system provides the muon identification in the pseudorapidity region up to ±4.6. Increasing the outer diameter of the forward toroid disks can compensate decreasing the bending power of the muon system at the pseudorapidity of ±2.5.

### Conclusions

The study investigates the idea of the superconducting solenoid magnet with the steel yoke for the hadron Future Circular Collider with a center-of-mass energy of 100 TeV. The parameters of the solenoid coil and the steel yoke seem to be reasonable. The magnet provides the required free bore of 12 m diameter and the central magnetic flux density of 6 T. The magnetic flux density distribution allows measuring the charged particle momenta in the pseudorapidity interval of ±3.5, and the conventional forward toroids increase the region for the muon identification up to the pseudorapidity of ±4.6. To increase the region of the charged particle measurements to the large pseudorapidity values, the dipole magnet with the bending power of at least 5 T·m should be considered between the end-cap disks and the forward muon toroids, and opening the end-cap inner cone should be decreased at the least to the pseudorapidity value of ±3.